\documentclass[aps,twocolumn,floats,superscriptaddress,showpacs]{revtex4}

\usepackage{graphicx}

\begin{document}

\title{Many-Body Approximation Scheme Beyond GW}

\author{Ping Sun} 
\affiliation{Department of Physics and
Astronomy, Rutgers University, Piscataway, NJ 08854-8019}
\author{Gabriel Kotliar}
\affiliation{Department of Physics and Astronomy,
Rutgers University, Piscataway, NJ 08854-8019}

\date{\today}

\begin{abstract}
We explore the combination of the extended dynamical mean field theory
(EDMFT) with the GW approximation (GWA); the former sums the local
contributions to the self-energies to infinite order in closed form
and the latter handles the non-local ones to lowest order. We
investigate the different levels of self-consistency that can be
implemented within this method by comparing to the exact QMC solution
of a finite-size model Hamiltonian. We find that using the EDMFT
solution for the local self-energies as input to the GWA for the
non-local self-energies gives the best result.
\end{abstract}


\maketitle

{\it Introduction.}-- The GW approximation (GWA) \cite{hedin} is one
of the most successful methods to describe electronic structure of
weakly correlated materials. It includes the lowest order perturbative
corrections in the screened Coulomb interaction to the electron
self-energy. In real space-time, the electron self-energy is given by
the product ``GW'', where ``G'' represents the electron Green's
function and ``W'' the RPA screened Coulomb interaction. The GWA has
been successfully applied to the calculations of quasi-particle
spectra of semiconductors and insulators.
\cite{strinati} (For recent reviews, see
Ref.\cite{ferdi1}) It describes well the experimentally
observed energy gaps in semiconductors
\cite{ferdi1,louie2,schilf1,eguiluz1,louie3,schilf2}.

The GWA self-energies were first calculated as a one-shot perturbation
by using the unperturbed LDA Green's functions. \cite{strinati}
However, as we will show in this paper, the one-shot GWA breaks down
as soon as the correlations are moderately strong, as expected for a
non-self-consistent leading order approximation. The problem should be
solved by including higher order corrections to extend the scope of
the method. This is the main motivation of our work. The Baym-Kadanoff
formulation \cite{baym} of the GWA, which makes the method
automatically conserving, requires to evaluate the self-energy, the
polarization bubble, and the Green's function self-consistently
\cite{hedin}. While this has been successfully implemented
\cite{groot} and shows improvement on the values of the total energy,
the self-consistent GWA brings up a debate over whether the fully
self-consistency improves the spectra.
\cite{eguiluz1,louie3,schilf2,ferdi2,godby,holm2,eguiluz2}. This is
due to the fact that the vertex correction, which is omitted in GWA,
shows a tendency to cancel the self-energy insertion when the full
Green's function is used in calculating the exchange self-energy
\cite{sernelius,mahan,holm1}. As a result the one-shot GWA or
partial-self-consistent ones (in which, {\it e.g.}, one fixes the
``W'' obtained from the LDA input while solving ``G''
self-consistently) are favored sometimes
\cite{ferdi2,godby,holm2}. This issue entails the usage of higher
order self-energy diagrams and provides us a second motivation for
improving systematically the approximation beyond the leading order.

In this paper, we extend the GWA by using the extended dynamical mean
field theory (EDMFT) \cite{edmft1,edmft3,edmft4,ping} which treats all
the local polarization, self-energy and vertex corrections in closed
form, following the ideas proposed in
Refs.\cite{edmft4,ping,sulki}. According to these approaches, the
local polarization and self-energy should be solved by EDMFT
non-perturbatively while the non-local ones via GWA. The cancellation
of the self-energy insertion and the vertex correction is carried out
{\it locally} to infinite order within EDMFT. However, the same
ambiguity of implementation (with or without self-consistency) remains
in the non-local GWA part. This question can not be addressed without
benchmarking the different schemes.

{\it Main Results.}-- We compared and contrasted the different
implementations of combining EDMFT with GWA. Our numerical calculation
suggests that the schemes using the EDMFT solution of the local
self-energies as input to the GWA for the non-local self-energies give
the best result. In other words, it is favorable {\it not} to allow
the feedback of the non-local GWA on the local EDMFT self-energies due
to different nature of the two methods, GWA being perturbative and
EDMFT non-perturbative. On the other hand, when using the EDMFT result
as input, it makes little difference whether one performs GWA with a
single shot or partial self-consistency (it is partial because the
local polarization and self-energy from EDMFT are fixed). This
supports the picture \cite{ping,sulki} that, in a correlated phase far
away from a phase transition, the temporal correlations reflected
through the local polarization and self-energy are dominant and should
be treated non-perturbatively, while the spatial correlations are
weaker and can be handled perturbatively.

\begin{figure}[ht]
\includegraphics[width=7.0cm]{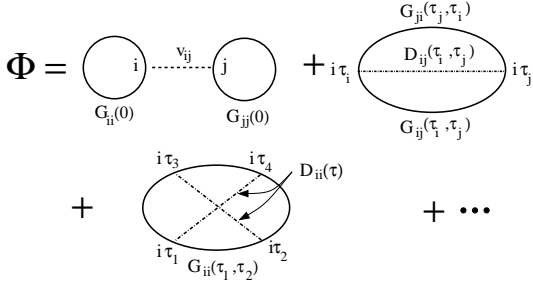}
\caption{The potential $\Phi$ of the Baym-Kadanoff functional
\cite{baym,edmft3} for (EDMFT+GW)$_{SC}$. (EDMFT)$_{SC}$ is obtained
by restricting the exchange diagram (the second on the r.h.s.) to be
local in space, $i=j$. The first line by itself represents the
standard (GW)$_{SC}$ scheme. In this formulation, the boson Green's
function $D$ describes the screened interaction. The boson self-energy
plays a similar role as the electron-hole bubble in GWA.}
\label{fig-edgw}
\end{figure}

\begin{figure}[ht]
\includegraphics[width=7.0cm]{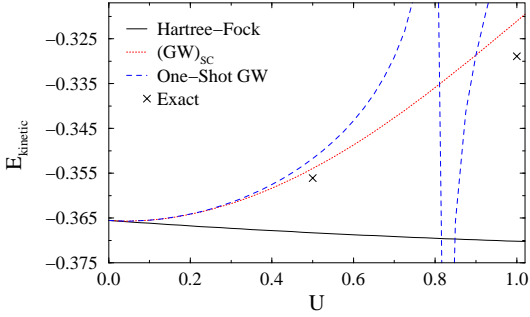}
\caption{In our model system, the one-shot GWA, using the Hartree-Fock
result as input, breaks down at $U \simeq 0.825$. This is an
instability against the formation of charge density wave at
$(\pi,\pi)$ \cite{ping1}. The other results are plotted as
references.}
\label{fig-g0w0}
\end{figure}

\begin{figure}[ht]
\includegraphics[width=6.3cm]{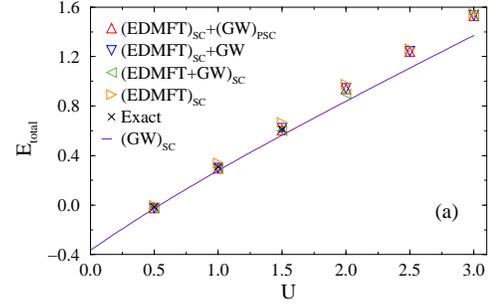}
\includegraphics[width=6.3cm]{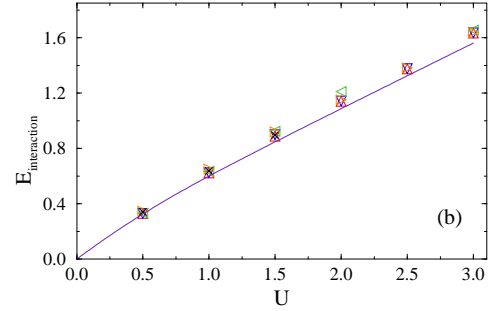}
\includegraphics[width=6.3cm]{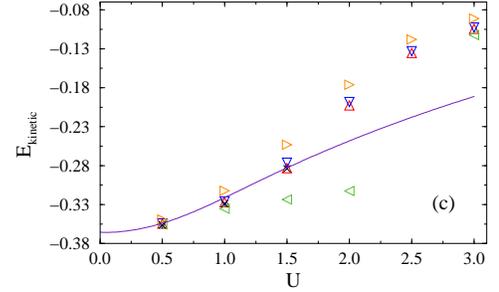}
\includegraphics[width=6.3cm]{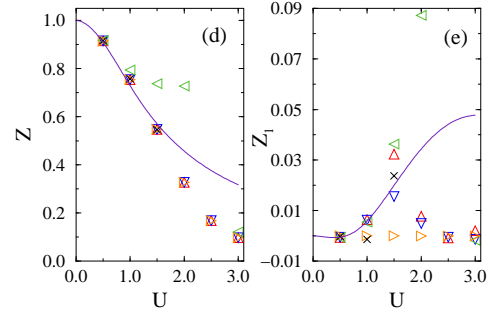}
\caption{The energies, the quasi-particle residue ($Z$), and the
non-local contribution to the self-energy ($Z_1$) as defined in the
text. The same symbol scheme defined in (a) applies to all the
diagrams. $Z$ is the same for (EDMFT)$_{SC}$+(GW)$_{PSC}$,
(EDMFT)$_{SC}$+GW, and (EDMFT)$_{SC}$, since the same local
self-energy is used. For $Z_1$, which reflects the spatial extension
of the self-energy, we are comparing numbers at least one order
smaller than those for other quantities.}
\label{fig-result1}
\end{figure}

\begin{figure}[ht]
\includegraphics[width=7.0cm]{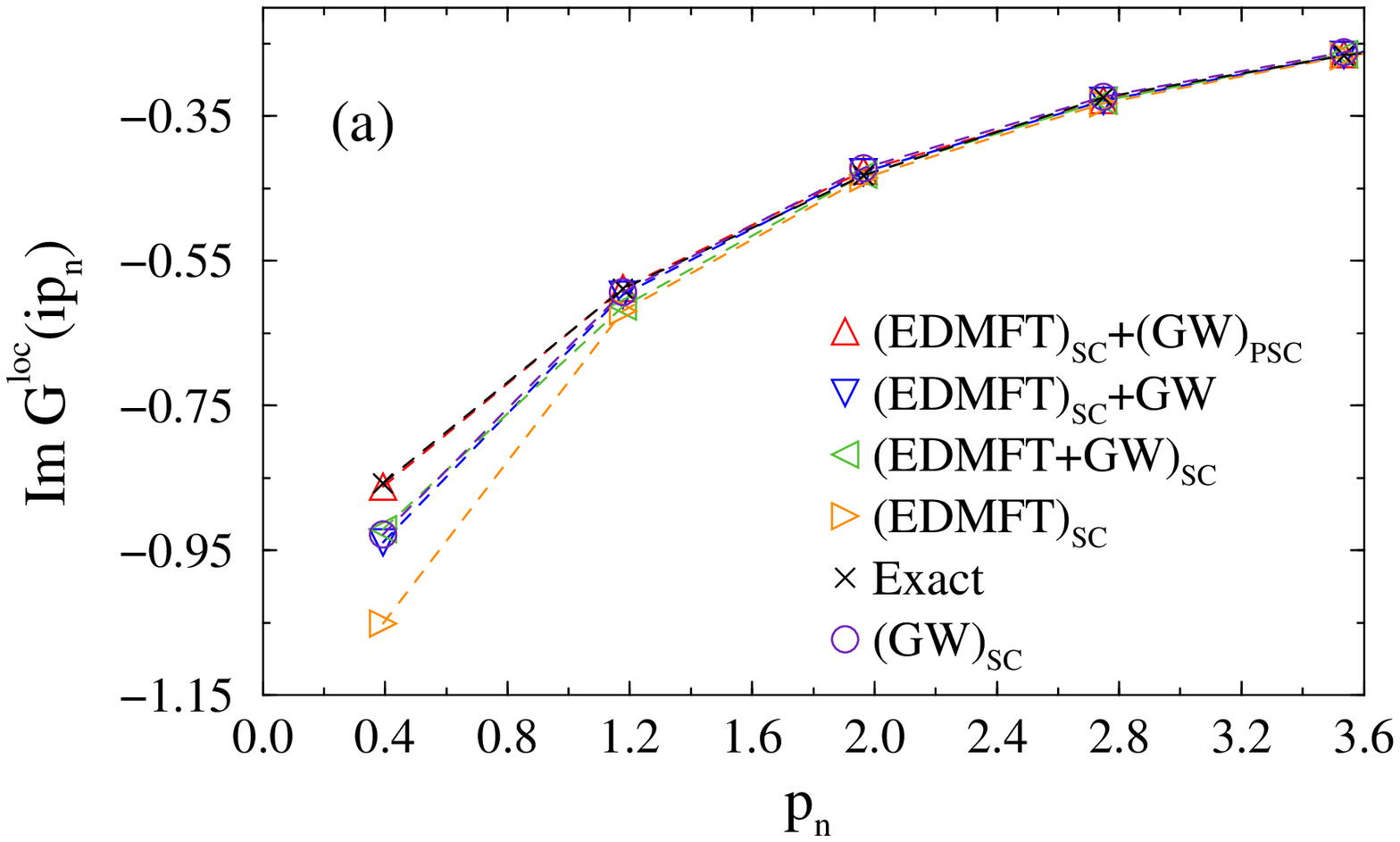}
\includegraphics[width=7.0cm]{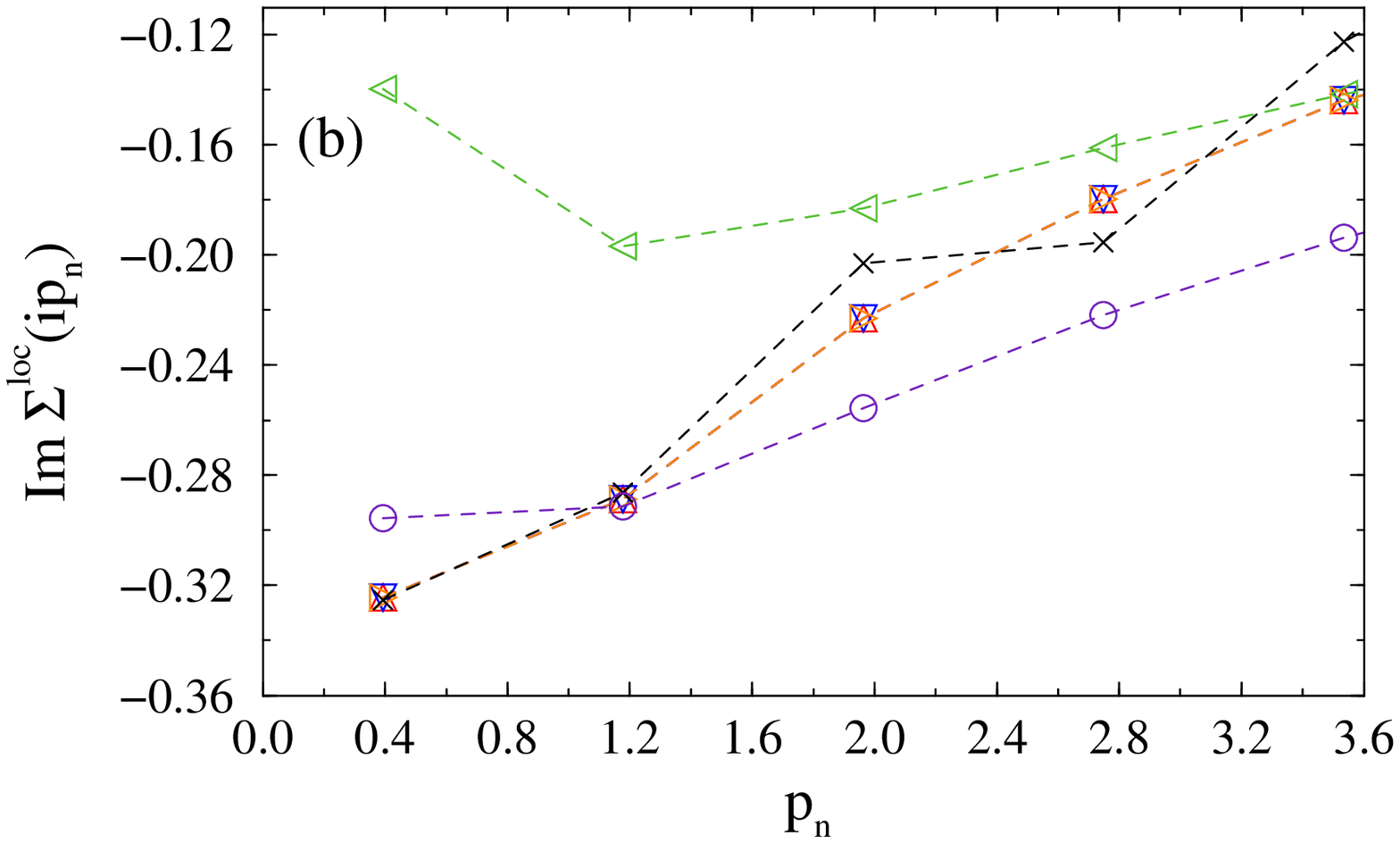}
\caption{The local electron Green's function and self-energy at
$U=1.5$. The same symbol scheme defined in (a) applies to (b). The
lines connecting the points are guides to the eye. The fermion
Matsubara frequency $p_n=(\pi/\beta)(2n+1)$.}
\label{fig-result2}
\end{figure}

{\it Lattice Model.}-- In comparing different GW approaches, many of
which apply to realistic materials, one always encounters the problem
that there are as many important differences in the implementations as
in the GWA methodologies themselves. These include the implementations
of GWA in imaginary time or frequency space, the choice and
construction of the local basis set, and the further approximations
like the plasmon-pole approximation \cite{ferdi1}. In order to compare
the many-body schemes without additional complications, it is desired
to employ simple model systems on the lattice
\cite{ferdi2,godby3}. Following the same strategy, we study the
following generalized Hubbard model:

\begin{equation}
\label{eq-01}
  \hat{H}=-\frac{1}{2}\sum_{ij,\sigma} t_{ij}
  (\hat{c}_{i\sigma}^{\dagger}\hat{c}_{j\sigma}
  +\hat{c}_{j\sigma}^{\dagger}\hat{c}_{i\sigma})
  +\frac{1}{2}\sum_{ij} \hat{n}_{i} v_{ij} \hat{n}_{j}
\end{equation}

\noindent where $\hat{c}_{i\sigma}$ ($\hat{c}_{i\sigma}^{\dagger}$)
annihilates (creates) an electron of spin $\sigma$
($=\uparrow,\downarrow$) at the lattice site $i$. $\hat{n}_{i}$ is the
electron density. We only consider the paramagnetic phase.

{\it Approximation Schemes.}-- The self-consistent schemes are
summerized in Fig.\ref{fig-edgw}. [we use ($\cdots$)$_{SC}$ for a
self-consistent (SC) loop and ($\cdots$)$_{PSC}$ a partially SC loop.]
The full functional in Fig.\ref{fig-edgw} can be separated into local
and non-local parts, which are functionals of the local and non-local
Green's functions, respectively:
$\Phi=\Phi_{EDMFT}[G_{Local},D_{Local}]+\Phi_{Non-Local \;
GW}[G_{Non-Local},D_{Non-Local}]$. The local (non-local)
self-energies, which are obtained from the functional derivatives of
$\Phi_{EDMFT}$ ($\Phi_{Non-Local \;GW}$), are functions of the local
(non-local) parts of the Green's functions only:
$\Sigma_{EDMFT}=\Sigma_{EDMFT}(G_{Local},D_{Local})$ and
$\Sigma_{Non-Local \; GW}=\Sigma_{Non-Local \;
GW}(G_{Non-Local},D_{Non-Local})$ for the electron,
$\Pi_{EDMFT}=\Pi_{EDMFT}(G_{Local},D_{Local})$ and $\Pi_{Non-Local \;
GW}=\Pi_{Non-Local \; GW}(G_{Non-Local})$ for the boson.

The approximation schemes we tested in combining EDMFT with GW are:
(i) The fully self-consistent (EDMFT+GW)$_{SC}$ solves the full Dyson
equations for $G=G_{Local}+G_{Non-Local}$ and
$D=D_{Local}+D_{Non-Local}$:
$G=[G_0^{-1}-\Sigma_{EDMFT}-\Sigma_{Non-Local \; GW}]^{-1}$,
$D=[D_0^{-1}-\Pi_{EDMFT}-\Pi_{Non-Local \; GW}]^{-1}$, with $G_0$ and
$D_0$ the free electron and boson Green's functions on the
lattice. The non-self-consistent schemes begin with the solution of
the EDMFT which solves
$G_{Local}=[G_0^{-1}-\Sigma_{EDMFT}]^{-1}_{Local}$ and
$D_{Local}=[D_0^{-1}-\Pi_{EDMFT}]^{-1}_{Local}$. (ii)
(EDMFT)$_{SC}$+GW uses the local EDMFT self-energies to calculate the
non-local Green's functions, $G_{EDMFT,
Non-Local}=[G_0^{-1}-\Sigma_{EDMFT}]^{-1}_{Non-Local}$ and $D_{EDMFT,
Non-Local}=[D_0^{-1}-\Pi_{EDMFT}]^{-1}_{Non-Local}$. It obtains in one
shot the estimation for $\Sigma_{Non-Local \; GW}=\Sigma_{Non-Local \;
GW}(G_{EDMFT, Non-Local},D_{EDMFT, Non-Local})$ and $\Pi_{Non-Local \;
GW}=\Pi_{Non-Local \; GW}(G_{EDMFT, Non-Local})$. (iii)
(EDMFT)$_{SC}$+(GW)$_{PSC}$ solves for $G_{Non-Local}$ and
$D_{Non-Local}$ self-consistently from
$G_{Non-Local}=[G_0^{-1}-\Sigma_{EDMFT}-\Sigma_{Non-Local \;
GW}]^{-1}_{Non-Local}$ and
$D_{Non-Local}=[D_0^{-1}-\Pi_{EDMFT}-\Pi_{Non-Local \;
GW}]^{-1}_{Non-Local}$ with $\Sigma_{EDMFT}$ and $\Pi_{EDMFT}$ fixed.

{\it Benchmark.}-- To compare and contrast the approximation schemes,
we perform a benchmark calculation using the model (\ref{eq-01}) on a
$4\times 4$, 2D square lattice with periodic boundary condition. We
use, for the free electron dispersion, $
\epsilon_{\vec{k}}=-(1/2)\left[\cos(k_x)+\cos(k_y)\right]$. The
half-bandwidth is taken as the energy unity. The interaction is given
by $ v_{\vec{k}}=U+2V\left[ \cos(k_x)+\cos(k_y)\right]$. We study the
half-filling case where the strongest correlation shows up. We fix the
inverse temperature $\beta=8.0$ and the ratio $V/U=0.25$. We vary $U$
from $0.0$ to $3.0$ for the approximation schemes. We benchmark the
calculation at $U=0.5, 1.0, 1.5$ by using direct QMC calculation via
the Hirsch-Fye algorithm \cite{hirsch}. In all the results we are
going to present, the major error of the calculation comes from the
QMC part, in both the exact solution and the EDMFT impurity solver.
In the latter we solve the EDMFT electron-boson impurity problem by a
hybridized Monte Carlo method \cite{buendia,motome,ping,ping1} which
employs an additional continuous auxiliary field. \cite{note}

At the given V-U ratio, a charge density wave instability at wave
vector $(\pi,\pi)$ is present in the one-shot GW. The breakdown is
shown in Fig. \ref{fig-g0w0}. In the exact solution, however, no
charge or spin instability is observed up to $U=1.5$. To compare the
different schemes on the same footing, we restrict ourselves to the
paramagnetic phase. Depending on the different schemes, this may mean
to studying a paramagnetic metastable solution when the possible
instability appears. We find it allowed for (GW)$_{SC}$ up to the
largest $U$ ($=3.0$) we studied and same for the EDMFT related schemes
we described, except for (EDMFT+GW)$_{SC}$ in which the
self-consistent solution does not exist at $U=2.5$.

Our main results are presented in Figs. \ref{fig-result1} and
\ref{fig-result2} in which the results from the exact QMC and the
(GW)$_{SC}$ are also plotted. The definitions of the plotted
quantities are as follows: The $E_{kinetic}$ and $E_{interaction}$ are
the hopping and interaction energies per site, calculated via the
Galitskii-Migdal formula \cite{galitskii}. $E_{total}=E_{kinetic}+
E_{interaction}$. The local electron self-energy of (EDMFT)$_{SC}$ is
obtained directly from the EDMFT solution while for those with spatial
extension, $\Sigma^{loc}(ip_n)=(1/N) \sum_{\vec{k}}
\Sigma(\vec{k},ip_n)$, with $N=4 \times 4$. The local Green's function
$G^{loc}(ip_n)=(1/N) \sum_{\vec{k}} \{[G_0(\vec{k},ip_n)]^{-1}-
\Sigma(\vec{k},ip_n)\}^{-1}$. The quasi-particle residue $Z=[1-{\rm
Im} \; \Sigma^{loc}(ip_0)/p_0 ]^{-1}$. We measure the spatial
extension of the electron-self-energy by $ Z_1=(1/N) \sum_{\vec{k}}
{\rm Im} \{ [\Sigma(\vec{k},ip_1)-\Sigma(\vec{k},ip_0)]/( p_1-p_0)\}
\cos k_x$.

From the {\it overall} features of the results, we see: (i) The
(EDMFT)$_{SC}$+(GW)$_{PSC}$ gives the best results up to the largest
benchmarked $U$ at $1.5$ times the half-bandwidth. (ii) The difference
between (EDMFT)$_{SC}$+(GW)$_{PSC}$ and (EDMFT)$_{SC}$+GW is small
quantitatively, since the local self-energies from (EDMFT)$_{SC}$ is
dominant. (iii) (EDMFT+GW)$_{SC}$ is not a good scheme in the sense
that it sees an artificial charge density instability around $U\sim
2.5$, where no self-consistent solution is found. Even before reaching
that regime, the (EDMFT+GW)$_{SC}$ shows significant deviations from
the exact solution. (iv) From the plottings of $Z$ and $Z_1$ in
Fig. \ref{fig-result1}, we see that (GW)$_{SC}$ misses the crossover
to localization at large $U$, while the EDMFT based calculations
correctly capture this feature. Actually, all the EDMFT related
schemes give results close to each other at $U=3.0$ since the spatial
extensions of the self-energies are no longer important. Our results
show that, with moderate and strong correlations, one needs to include
higher order contributions beyond GWA. The schemes,
(EDMFT)$_{SC}$+(GW)$_{PSC}$ and (EDMFT)$_{SC}$+GW, offer a reliable
solution towards this direction.

{\it Conclusion.}-- To summarize, we presented a many-body scheme
which handles the local self-energies non-perturbatively via EDMFT and
the non-local ones perturbatively via GWA. As an improvement over the
leading order GWA, the new scheme better captures the effects of
correlation. We described several implementations, which perform
self-consistency at different levels, and benchmarked them by
comparing with the exact solution of a finite-size model system. We
found that (EDMFT)$_{SC}$+(GW)$_{PSC}$ and (EDMFT)$_{SC}$+GW gave very
close results. For the model we studied, (EDMFT)$_{SC}$+(GW)$_{PSC}$
gave the best result.

Finally, we should point out that, similar to the self-consistent GWA
\cite{holm2,mahan}, our scheme has the problem that the polarization
does not have the proper asymptotic behavior in the long wave length
limit [$\lim_{k\rightarrow 0} P(k,\omega) \propto (k/\omega)^2$].
Since the schemes combining EDMFT with GWA include local
vertex corrections non-perturbatively, this violation is less severe
than that in GWA. \cite{ping1,note2}

{\it Acknowledgements.}-- This research was supported by NSF under
Grant No. DMR-0096462 and by the Center for Materials Theory at
Rutgers University. The authors would like to thank A. Lichtenstein
based on whose multiband QMC program part of the exact QMC calculation
was performed. P.S. would like to thank the helpful discussions with
H. Jeschke, W. Ku, C. Marianetti, V. Oudovenko, S. Savrasov, R. Scalettar,
M. Schilfgaarde, and S. Zhang.


\begin{thebibliography}{99}
\bibitem[28]{note} In QMC, we use 32 time slices. For each EDMFT
iteration, we perform $10^6$ QMC sweeps and usually EDMFT converges
after about 10 iterations. In the case of the exact solution, due to
the sign problem, the effective number of QMC sweeps is reduced from
the actual number $\sim 5\times 10^4$. In the worst case at $U=1.5$
the effective sweep number becomes $\sim 10^4$. To proceed beyond
$U=1.5$, one needs to use approximated algorithms other than the exact
QMC to overcome the sign problem.
\bibitem[30]{note2} At $U=1.5$, we find the polarizations
  $P(k=0,i\omega_n)$ at $n=32$ (so $\omega_n=25.13$) are: -1.88E-3 for
  (GW)$_{SC}$, -4.6E-4 for (EDMFT)$_{SC}$, -1.3E-4 for
  (EDMFT+GW)$_{SC}$, -6E-5 for (EDMFT)$_{SC}$+GW, and 8E-5 for
  (EDMFT)$_{SC}$+(GW)$_{PSC}$. In all the cases we studied, the GWA
  always results in a much bigger violation than the EDMFT related
  methods, especially those obtained from combining EDMFT with GWA.
\end{thebibliography}
\end{document}